\documentclass{Interspeech}

\interspeechcameraready

\title{WhisperD: Dementia Speech Recognition and Filler Word Detection with Whisper}

\author{Emmanuel}{Akinrintoyo}
\author{Nadine}{Abdelhalim}
\author{Nicole}{Salomons}

\affiliation[nocounter]{I-X and Department of Computing}{Imperial College London}{UK}
\email{[e.akinrintoyo23, nadine.abdelhalim23, n.salomons]@imperial.ac.uk}
\keywords{Dementia, Whisper, ASR, Filler words, Speech-to-text, Speech recognition, Disfluency, Fine-tuning}

\usepackage{comment}
\usepackage{array}
\usepackage{pgfplots}
\pgfplotsset{compat=1.17}
\usepackage{graphicx} 
\usepackage{caption}  
\usepackage{booktabs}
\usepackage{xcolor}
\usepackage{dirtytalk}
\usepackage{pgfplotstable}

\begin{document}

\maketitle

\begin{abstract}
Whisper fails to correctly transcribe dementia speech because persons with dementia (PwDs) often exhibit irregular speech patterns and disfluencies such as pauses, repetitions, and fragmented sentences. It was trained on standard speech and may have had little or no exposure to dementia-affected speech. However, correct transcription is vital for dementia speech for cost-effective diagnosis and the development of assistive technology. In this work, we fine-tune Whisper with the open-source dementia speech dataset (DementiaBank) and our in-house dataset to improve its word error rate (WER). The fine-tuning also includes filler words to ascertain the filler inclusion rate (FIR) and F1 score. The fine-tuned models significantly outperformed the off-the-shelf models. The medium-sized model achieved a WER of 0.24, outperforming previous work. Similarly, there was a notable generalisability to unseen data and speech patterns.
\end{abstract}

\section{INTRODUCTION}


Dementia is an umbrella term that describes various symptoms that impact memory, thinking, communication, and social abilities~\cite{geldmacher1996evaluation}. Persons with dementia (PwDs) experience a cognitive decline that impacts their ability to communicate and use language effectively~\cite{geldmacher1996evaluation}. The decline stems from neurodegenerative changes in the brain that affect aspects of language such as speaking. PwDs may struggle to find the right words or form coherent sentences, often exhibiting incorrect or irregular grammar~\cite{guinn2012language}. PwDs speak with nuances that includes an increased use of disfluencies, such as pauses, irregular pitch and volume of the voice, and filler words~\cite{guinn2012language}.

Filler words are sounds or unnecessary words that do not contribute to the meaning of a sentence but are used by people to fill pauses or give themselves time to think~\cite{duvall2014exploring, zhu2022filler}. Filler words (such as \textit{uh} and \textit{um}) are prevalent in speech impairments associated with conditions such as dementia~\cite{guinn2012language}. Patterns and use of fillers by PwDs are vital markers of cognitive impairment~\cite{bortfeld2001disfluency, guinn2012language}. Fillers are used when a PwD is experiencing cognitive load or strain, is unsure of what to say, or is trying to process their response. Hence, fillers contain rich information that can be used to detect dementia or track its progression~\cite{woszczyk2024prosody}. For instance, some previous research has shown the effectiveness of dementia detection from transcripts that included repeated speech portions and filler words~\cite{matovsevic2022accurate}. 

Adaptive and personalised responses can be provided by smart devices, such as conversational agents or robots through filler detection. By analysing fillers, a system can identify common patterns of hesitation in the speech of a PwD. This can then be used to deliver a more tailored and supportive interaction. 

Furthermore, fillers can also help detect symptoms of disorientation during a conversation with a PwD~\cite{mirheidari2018detecting}. Disorientation is often reflected in an increased duration, frequency, and irregular usage of fillers. When a PwD is disorientated, they can struggle to process information or recall details, thus leading to an increase in filler words~\cite{ovchinnikova2017lexical}. 

In addition, the quality of speech of PwDs is degraded as the syndrome progresses~\cite{mirheidari2018detecting, guinn2012language}. Thus, a longitudinal study of disfluencies and filler patterns over a period can be combined with other cognitive indicators to identify the progression of dementia. This can aid caregivers and dementia professionals in making proactive adjustments for a PwD’s cognitive care plan~\cite{zolnoori2023adscreen}.

Automatic Speech Recognition (ASR) systems such as OpenAI's Whisper~\cite{radford2023robust} offer cost-effective solutions for dementia speech understanding, including the ability to detect fillers. Despite its impressive performance for speech transcription, previous research has shown that Whisper has some limitations. Similar to previous ASR solutions, it suffers performance degradation with pathological speech~\cite{sanguedolce2024whisper}. Notably, Whisper is designed to generate clean output transcriptions. It mostly excludes filler words from its outputs. 


\subsection{Prior Work}

Previous work has explored various solutions for understanding dementia speech. This includes the conformer-based speech recognition system by Wang et al.~\cite{wang2022conformer}. The system's accuracy was assessed with the WER (word error rate). It achieved a reduction of 13.6\% absolute (34.8\% relative) to the baseline while also achieving the best-reported accuracy of 91.2\% for Alzheimer's detection. Others have used Whisper for dementia detection and classification~\cite{10825055, 10448004} and have recorded significant performances compared to previous work with F1 scores of 86.42\% and 84.50\%.

Other research has explored systems for filler detection and classification specifically for the dementia population~\cite{rohanian2021alzheimer, nasreen2021alzheimer, zolnoori2023adscreen}. Previous work such by Soleimani et al.~\cite{soleimani2024impact} found that dementia detection is feasible by encoding filler words (such as \textit{uh} and \textit{um}) in text-based language models.

Sanguedolce et al.~\cite{sanguedolce2024whisper} demonstrated Whisper's potential to achieve excellent performance when fine-tuned for pathological speech with fillers included. The authors fine-tuned Whisper for persons with stroke (PwS) and obtained a WER of 21.5\% on the AphasiaBank dataset, surpassing previous models. Previous research by Wagner et al.~\cite{wagner2024crisperwhisper} examined Whisper's fine-tuning for filler word inclusion. A state-of-the-art performance was achieved by adjusting the model tokeniser with F1 scores greater than 0.92 for collar values between 0.5 and 0.6 seconds on the PodcastFillers Corpus~\cite{zhu2022filler}. We adopt a similar approach to Wagner et al.~\cite{wagner2024crisperwhisper} with Whisper by adjusting the tokeniser to train it on dementia speech.

In this work, we explore Whisper's performance when fine-tuned on the speech of PwDs for improved WER. Unlike previous work, we utilise the DementiaBank's Pitt and Kempler datasets with our internal dementia speech dataset, achieving state-of-the-art performance. We modify Whisper's tokeniser to include filler words (\textit{uh} and \textit{um}). Filler words are included in the dataset for the model's training and inference. In the subsequent sections, we present the methodology adopted with the findings and evaluation of the model.

\section{METHODOLOGY}

\subsection{Dataset}
The DementiaBank dataset was utilised to fine-tune Whisper. This comprised of the Pitt~\cite{pitt}, and Kempler~\cite{Kempler} datasets. The Pitt dataset consists of 306 PwDs in a longitudinal study. The participants completed tasks such as describing the Boston Cookie Theft stimulus photo~\cite{cookie_theft}, word fluency, story recall and sentence construction. Similarly, the Kempler dataset contains descriptions of the Cookie Theft picture and some conversations between the six participants and the investigators. The audio files were transcribed manually using the CHAT (Codes for the Human Analysis of Transcripts) protocol~\cite{macwhinney2014childes}.

In addition, we used our in-house \textit{CONNECT} dataset collected from our robot-based individual cognitive stimulation therapy (iCST)~\cite{akinrintoyo2024co, akinrinE} sessions with PwDs. iCST is a non-pharmacological personalised intervention that aims to improve cognitive function, quality of life, and social participation of people with mild to moderate dementia~\cite{yates2015development}. Five PwDs ($\bar{x}$ = 83.2 years) completed one-to-one 20-minutes sessions with a Misty robot~\cite{srinivasan2019misty} on two iCST activities. This included famous places and common sayings activities in which PwDs discussed about landmarks and popular proverbs and their meanings. The audio samples were transcribed and verified by two of the researchers.

\subsection{Dataset Pre-Preprocessing}
Sentence-level segmentation was adopted as a pre-processing step to divide continuous text into individual sentences. This was necessary to ensure that each audio segment remains within the optimal length constraint of 30 seconds. This was influenced by the design limitations of Whisper. Whisper is not optimised for processing audio inputs exceeding 30 seconds in a single pass. Hence, longer recordings must be segmented into smaller chunks to preserve the natural structure of the speech.

The interviewers' speech was removed from the audio during the segmentation, leaving only the speech of PwDs in the dataset. The transcriptions were cleaned to ensure text and language consistency. This involved deleting text markers such as \say{xxx}, which were used to indicate segments where the speech of a PwD could not be deciphered. It also included normalising informal or colloquial words such as \say{hafta} to \say{have to}. 

A similar training approach to the original methodology for Whisper is adopted by incorporating speech and non-speech segments. This is to enhance WhisperD's ability to accurately detect and process dementia speech. It is crucial to improve voice activity detection (VAD), which is vital to differentiate between spoken and silent periods. This is particularly beneficial for dementia speech, where PwDs often speak with frequent pauses, extended silences, and irregular speech rhythms. Hence, the inclusion of non-speech segments reduces the likelihood of misclassifying silence as speech or vice versa. Moreover, it increases WhisperD's robustness to real-world scenarios where speech is interspersed with silence, background noise, or other non-verbal sounds.

Furthermore, we excluded audio files with a duration below one second from the dataset to avoid potential processing issues for Whisper. Hence, all audio files in the dataset have a length between 1 and 30 seconds, inclusive. The audio files were converted to \textit{.wav} format to ensure suitability with the model. A resampling size of 16,000 Hz was used for all audio files, similar to the off-the-shelf models.

There were a total of 11,397 audio samples after cleaning and sentence-level segmentation. The speech dataset was a total of 11.39 hours, with large portions of silence removed from the original datasets. The segmentation accuracy was verified by the authors. This was done by using the Whisper Large variant to transcribe each audio sample for a comparison with the original transcripts obtained from the CHAT (Codes for the Human Analysis of Transcripts) files. Audio files with a similarity index below 70\% were then carefully reviewed to correct any transcription errors. 70\% was chosen as a threshold to balance sensitivity and efficiency. Audio files with speech that were completely inaudible were excluded from the dataset.

Disfluencies such as filler words (\textit{uh} and \textit{um}) are included in the dataset to ensure it learns to include them in its output. There were a total of 1699 \textit{uh} and 614 \textit{um} in the dataset. We found that \textit{uh} is more prevalent in dementia speech. Symbols and characters were excluded from the dataset, except apostrophes. Extra spaces were also removed for clean transcriptions.

\begin{table*}[h]
    \centering
    \renewcommand{\arraystretch}{1.3} 
    \resizebox{\textwidth}{!}{ 
    \begin{tabular}{>{\centering\arraybackslash}p{3.5cm} >{\centering\arraybackslash}p{2cm} >{\centering\arraybackslash}p{2cm} 
    >{\centering\arraybackslash}p{2cm} >{\centering\arraybackslash}p{2cm} >{\centering\arraybackslash}p{2cm} >{\centering\arraybackslash}p{2cm}}  
        \toprule
        \textbf{MODEL} & \multicolumn{3}{c}{\textbf{VALIDATION SET}} & \multicolumn{3}{c}{\textbf{TEST SET}} \\ 
        \cmidrule(lr){2-4} \cmidrule(lr){5-7} 
        & \textbf{WER} & \textbf{FIR} & \textbf{F1} & \textbf{WER} & \textbf{FIR} & \textbf{F1} \\  
        \midrule
        Whisper-T  & 0.77 & 0.03 & 0.05 & 0.89 & 0.07 & 0.13 \\ 
        \textbf{WhisperD-T}  & \textbf{0.66} & \textbf{0.50} & \textbf{0.61} & \textbf{0.50} & \textbf{0.42} & \textbf{0.57} \\ 
        Whisper-B  & 0.81 & 0.06 & 0.10 & 0.81 & 0.06 & 0.11 \\ 
        \textbf{WhisperD-B}  & \textbf{0.39} & \textbf{0.35} & \textbf{0.49} & \textbf{0.43} & \textbf{0.33} & \textbf{0.46} \\ 
        Whisper-S & 0.45 & 0.05 & 0.08 & 0.48 & 0.05 & 0.11 \\ 
        \textbf{WhisperD-S} & \textbf{0.29} & \textbf{0.73} & \textbf{0.72} & \textbf{0.28} & \textbf{0.72} & \textbf{0.76} \\ 
        Whisper-M & 0.42 & 0.06 & 0.11 & 0.39 & 0.04 & 0.09 \\ 
        \textbf{WhisperD-M} & \textbf{0.25} & \textbf{0.64} & \textbf{0.71} & \textbf{0.24} & \textbf{0.70} & \textbf{0.73} \\ 
        \bottomrule    
    \end{tabular}
    }
    \caption{Performance Metrics for Whisper Models (Tiny, Base, Small, Medium) and their WhisperD (fine-tuned) Variants: Comparison of Word Error Rate (WER), and Filler Inclusion Rate (FIR), F1 Score.} 
    \label{tab:performance_metrics}
\end{table*}

\subsection{Evaluation Metrics}
WhisperD's performance was evaluated at the end of each epoch using the Word Error Rate (WER) metric. WER compares the overall accuracy of the ASR by comparing the predicted transcript with the reference transcript. WER is the sum of substitutions, insertions and deletions divided by the total number of words in the reference transcription. A lower WER indicates better performance. WER is calculated as:

\[
\text{WER} = \frac{S + D + I}{N}
\]
where:
\begin{itemize}
    \item \( S \) = Number of substitutions (incorrect words replacing correct ones)
    \item \( D \) = Number of deletions (words omitted from the reference)
    \item \( I \) = Number of insertions (extra words added by the model)
    \item \( N \) = Total number of words in the reference transcription
\end{itemize}

Furthermore, Filler Inclusion Rate (FIR) was used as a metric. FIR measures how accurately the ASR transcribes filler words (\textit{uh} and \textit{um}) compared to the reference transcript. This is calculated by dividing the number of correctly detected fillers by the total number of fillers. FIR is set to one (i.e., a perfect score) to avoid division by zero when there are no filler words in the reference text. 


Since FIR ignores false positives, the F1 score was computed to provide a fuller picture of the model’s performance. The F1 score is the harmonic mean of precision and recall. 



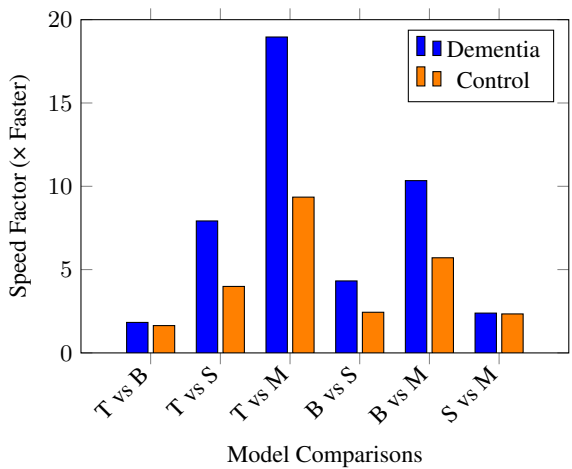
\begin{figure}
    \centering
    \begin{tikzpicture}
        \begin{axis}[
            ybar,
            symbolic x coords={T vs B, T vs S, T vs M, B vs S, B vs M, S vs M},
            xtick=data,
            ymin=0,
            ymax=20, 
            ylabel={Speed Factor (× Faster)},
            xlabel={Model Comparisons},
            width=\columnwidth,  
            height=6cm,
            font=\fontsize{9}{9}\selectfont,  
            xlabel style={font=\fontsize{9}{9}\selectfont}, 
            ylabel style={font=\fontsize{9}{9}\selectfont}, 
            bar width=8pt,
            enlarge x limits=0.2,
            xticklabel style={rotate=45, anchor=east},  
            legend style={at={(0.97,0.97)}, anchor=north east, draw=black, fill=white, font=\fontsize{9}{9}\selectfont} 
        ]
        \addplot [fill=blue] coordinates {
            (T vs B, 1.83) (T vs S, 7.92) (T vs M, 18.96) 
            (B vs S, 4.32) (B vs M, 10.34) (S vs M, 2.39)
        };
        \addlegendentry{Dementia}
        \addplot [fill=orange] coordinates {
            (T vs B, 1.64) (T vs S, 3.99) (T vs M, 9.35) 
            (B vs S, 2.44) (B vs M, 5.71) (S vs M, 2.34)
        };
        \addlegendentry{Control}
        \end{axis}
    \end{tikzpicture}
    \caption{Speed Comparison of Whisper Models (Tiny (T), Base (B), Small (S) and Medium (M)): Dementia vs Control Speech}
    \label{fig:speed_factor}
\end{figure}

\subsection{Whisper Training}
Whisper is a transformer-based architecture consisting of an encoder and a decoder with similar width and transformer blocks~\cite{radford2023robust}. An 80-channel log-magnitude Mel spectrogram is generated using 25-millisecond windows with a 10-millisecond stride. Whisper has different sizes: \textit{tiny} (Whisper-T, 39M parameters), \textit{base} (Whisper-B, 74M parameters), \textit{small} (Whisper-S, 244M parameters), \textit{medium} (Whisper-M, 769M parameters) and \textit{large} (Whisper-L, 1550M parameters). The tiny, base, small, and medium-sized models are examined in this work for developing variants of WhisperD. 

Pre-trained Whisper models were imported from Hugging Face. The PyTorch and Hugging Face Transformers libraries were used to fine-tune the standard models to develop WhisperD. We fine-tuned all the layers of Whisper to include both the encoder and decoder for optimal performance. The fine-tuning process involved adjusting a model's parameters to adapt it to the dementia speech dataset. Whisper's tokeniser was modified to include filler words (\textit{uh} and \textit{um}). This is to enhance the model's ability to recognise and encode these filler words.

The dataset was split with a ratio of 80-10-10\% for the training, validation, and test sets. The training and validation sets contain the same speakers. However, these speakers were excluded from the test set. The speakers in the test set were selected from the Pitt dataset. This was done to ensure that the model is not evaluated on the speech patterns it encountered in the training phase. This prevents data leakage and allows for a more accurate assessment of the model's ability to generalise to unseen speakers. This approach is reliable for measuring real-world performance. 

AdamW, an adaptive optimisation algorithm, was chosen to provide training stability and generalisation. A learning rate of 1.2e-5 was used with a weight decay of 0.01 to reduce overfitting. The scheduler was set with a cosine learning rate decay. It had a warm-up period of 650 steps. Gradient accumulation was used with a batch size of eight and an accumulation step of eight, hence simulating a batch size of 64. The models were trained for a maximum of five epochs, and the best-performing model was chosen. Cross-entropy loss function was applied at each step to optimise token prediction. The loss is averaged across the batch during training. 

\section{Evaluation and Results}

\subsection{Off-the-Shelf Models}
We performed initial assessments of the off-the-shelf Whisper models. A transcription test was performed with 100 audio files with the tiny, base, small, and medium models. This was performed using Intel's 13th Gen i7-1370P Processor 1900 \textit{MHz} with 14 Core(s). The analysis revealed that due to Whisper's sequence-to-sequence style, it often uses the context of the sentence or words preceding to match the words that align next. This was verified by examining Whisper's token prediction array. For instance, the highest probability sequence for a speech segment was \say{Water is the picture, just tell me everything}. However, Whisper's final output was \say{This is the picture, just tell me everything}. 

While the structured output ensures grammatical correctness and clarity, it can limit the performance in transcribing dementia speech. A PwD can make corrections while speaking. PwDs can often change their thoughts mid-sentence (referred to as thought derailment~\cite{jeronimo2018formal}) with hesitations and fragmented statements. These nuances may be lost when the model prioritises fluency over authenticity.


The test results (Figure \ref{tab:performance_metrics}) showed that the Whisper-T model was 1.83× faster than the Whisper-B, 7.92× faster than the Whisper-S, and 18.96× faster than the Whisper-M. Whisper-B was 4.32× faster than the Whisper-S model and 10.34× faster than Whisper-M. The Whisper-S model was 2.39× faster than the Whisper-M model. A similar test was performed with the control speech dataset. It revealed that Whisper-T was 1.64× faster than Whisper-B, 3.99× faster than Whisper-S, and 9.35× faster than Whisper-M. Whisper-B was 2.44× faster than Whisper-S and 5.71× faster than Whisper-M. Whisper-S was 2.34× faster than the Whisper-M. 

The results revealed that transcribing dementia speech significantly impacts the processing speed of Whisper, especially for larger models. This could be due to the higher disfluency, increased ambiguity, and more decoding passes needed to transcribe dementia speech. Whisper may have to make repeated attempts at recognition.

Since Whisper tends to omit filler words such as \textit{uh} and \textit{um} in its transcriptions, it often considers a filler word as part of the next word such that it mistranscribes the actual word. For instance, \say{uh distinct} is transcribed as \say{staked}. In addition, Whisper outputted nothing when the volume was low due to a PwD speaking with a low voice, and hallucinates when it fails to understand the speech of a PwD. Notably, it struggled with how PwDs pronounced words such as \say{bureau} which it transcribed as either \say{biro}, \say{b-roll}, \say{pure}, or \say{do you roll?}.

\subsection{WhisperD vs. Off-the-Shelf Models}
Table \ref{tab:performance_metrics} compares the performance of four off-the-shelf Whisper models (Whisper-T, Whisper-B, Whisper-S, and Whisper-M) with their fine-tuned versions (WhisperD variants). The results show that the WhisperD models significantly outperform their corresponding Whisper models. Notably, the performance remains constant across both the validation and test sets for the best-performing models. 

WhisperD's performance for the test set demonstrates its ability to generalise well to unseen data and speech patterns across the variants. It achieved a low WER for the test set, with a better WER performance than the validation sets for all sizes except WhisperD-B. The average WER reduction for WhisperD models compared to their corresponding Whisper models is 0.36 and 0.43 for the validation and test sets, respectively. WhisperD-B and WhisperD-T have the highest WER performance improvement (0.42 and 0.39) from the standard models for the validation and test sets, respectively. 

WhisperD models outperform Whisper models in recall and precision (measured by F1 score) for filler word detection. They detect far more filler words in comparison to Whisper models. WhisperD-S has the best performance across the validation and test sets for filler detection. While FIR is high for WhisperD models, the increase in F1 indicates that it is due to improved detection rather than excessive false positives. WhisperD is not just over-inserting filler words, it is identifying them correctly. However, the F1 scores recorded in this work are below previous work by Wagner et al.~\cite{wagner2024crisperwhisper} which examined filler detection on standard speech with the PodcastFillers dataset containing over 17,000 instances of both \textit{uh} and \textit{um} individually.

\section{Discussion}

The standard Whisper models were trained on 30-second audio clips. This provides sufficient context for Whisper, such that it uses previously generated text tokens to ensure coherence in its prediction. WhisperD was trained with shorter audio clips. Hence, WhisperD is optimised to handle fragmented speech patterns, which are common in clinical and diagnostic settings. This allows it to excel in scenarios where short, isolated speech segments must be transcribed accurately without extended context. This is beneficial for use cases such as clinical interviews, cognitive assessments, and robot-assisted speech interventions, where responses are often brief with little or no contextual continuity.

WhisperD was developed with the original DementiaBank datasets without noise filters applied, unlike those used for the dementia classification challenges. Hence, the reported results are indicative of the model's real-world performance. Importantly, this work demonstrates the potential to develop ASR systems with an improved understanding of dementia speech. This is notable for ASRs with multi-lingual capabilities.

WhisperD models can serve as a cost-effective screening tool for dementia, without tester variability. They can be used to process and analyse speech patterns to identify early signs of cognitive decline, and detect dementia with progression tracking. Prior work such as Weiner et al.~\cite{weiner2017manual} has demonstrated the effectiveness of ASRs in dementia detection pipeline. Notably, WhisperD demonstrates that a multilingual model has the potential to detect dementia speech patterns across various languages and dialects when fine-tuned, providing accessibility for diverse populations.

Overall, WhisperD-M and WhisperD-S are the best models, balancing low WER and high F1 scores. WhisperD-S performed better with its highest F1 score (0.76 on the test set). However, WhisperD-M might be better for general use with its lowest WER (0.24 on the test set). The WER achieved by WhisperD-M outperforms previous work in~\cite{ mirheidari2018detecting, pan2020improving, ye2021development, wang2022conformer, pan2025two}.


\section{Limitations}
A dataset of 11.39 hours is relatively short. Expanding the training dataset would likely yield better results in terms of WhisperD's ability to learn domain-specific patterns and acoustic variations. Techniques such as speed perturbation can be used to increase the size of the dataset. In addition, the inclusion of speech disfluency datasets such as PodcastFillers~\cite{zhu2022filler} may increase the performance of WhisperD for filler detection. 

The quality of the dataset also influenced the overall performance. A key challenge encountered was the presence of mumbled and unclear speech in some audio samples. There were some unintelligible words that could not be transcribed in the DementiaBank datasets since they were too ambiguous or inaudible for human annotators and researchers. However, WhisperD attempted to predict them using the surrounding context. While this is beneficial, it led to increased WER. This affected WhisperD's performance, not necessarily due to poor transcription ability, but as a result of discrepancies between human-annotated transcripts and model predictions.


\section{Conclusion and Future Work}
In this work, WhisperD was proposed as a specialised model for transcribing dementia speech. It outperformed the standard Whisper models, highlighting its immense potential to serve as a comprehensive solution for dementia speech understanding with the inclusion of filler words. It achieved state-of-the-art WER performance, with strong generalisability to unseen speech. This can significantly improve early diagnosis and accessibility to clinical assessments and assistive devices.

Future work will examine the performance of the Whisper Large variant. It has the potential to achieve higher transcription accuracy and better robustness to disordered speech patterns. However, given the increased model complexity and parameter count, it may require more extensive datasets to reach its full potential. Therefore, future work will expand the dataset, especially through the collection of speech data from PwDs through our robot-based iCST study.


\bibliographystyle{IEEEtran}
\bibliography{mybib}

\begin{thebibliography}{10}
\providecommand{\url}[1]{#1}
\csname url@samestyle\endcsname
\providecommand{\newblock}{\relax}
\providecommand{\bibinfo}[2]{#2}
\providecommand{\BIBentrySTDinterwordspacing}{\spaceskip=0pt\relax}
\providecommand{\BIBentryALTinterwordstretchfactor}{4}
\providecommand{\BIBentryALTinterwordspacing}{\spaceskip=\fontdimen2\font plus
\BIBentryALTinterwordstretchfactor\fontdimen3\font minus \fontdimen4\font\relax}
\providecommand{\BIBforeignlanguage}[2]{{%
\expandafter\ifx\csname l@#1\endcsname\relax
\typeout{** WARNING: IEEEtran.bst: No hyphenation pattern has been}%
\typeout{** loaded for the language `#1'. Using the pattern for}%
\typeout{** the default language instead.}%
\else
\language=\csname l@#1\endcsname
\fi
#2}}
\providecommand{\BIBdecl}{\relax}
\BIBdecl

\bibitem{geldmacher1996evaluation}
D.~S. Geldmacher and P.~J. Whitehouse, ``Evaluation of dementia,'' \emph{New England Journal of Medicine}, vol. 335, no.~5, pp. 330--336, 1996.

\bibitem{guinn2012language}
C.~I. Guinn and A.~Habash, ``Language analysis of speakers with dementia of the alzheimer’s type,'' in \emph{2012 AAAI fall symposium series}, 2012.

\bibitem{duvall2014exploring}
E.~Duvall, A.~Robbins, T.~Graham, and S.~Divett, ``Exploring filler words and their impact,'' \emph{Schwa. Language \& Linguistics}, vol.~11, pp. 35--49, 2014.

\bibitem{zhu2022filler}
G.~Zhu, J.-P. Caceres, and J.~Salamon, ``Filler word detection and classification: A dataset and benchmark,'' in \emph{Proc. Interspeech 2022}, 2022.

\bibitem{bortfeld2001disfluency}
H.~Bortfeld, S.~D. Leon, J.~E. Bloom, M.~F. Schober, and S.~E. Brennan, ``Disfluency rates in conversation: Effects of age, relationship, topic, role, and gender,'' \emph{Language and speech}, vol.~44, no.~2, pp. 123--147, 2001.

\bibitem{woszczyk2024prosody}
D.~Woszczyk, R.~Aloufi, and S.~Demetriou, ``Prosody-driven privacy-preserving dementia detection,'' in \emph{Proc. Interspeech 2024}, 2024.

\bibitem{matovsevic2022accurate}
L.~Mato{\v{s}}evi{\'c} and A.~Jovi{\'c}, ``Accurate detection of dementia from speech transcripts using {Roberta} model,'' in \emph{2022 45th Jubilee International Convention on Information, Communication and Electronic Technology (MIPRO)}.\hskip 1em plus 0.5em minus 0.4em\relax IEEE, 2022, pp. 1478--1484.

\bibitem{mirheidari2018detecting}
B.~Mirheidari, ``Detecting early signs of dementia in conversation,'' Ph.D. dissertation, University of Sheffield, 2018.

\bibitem{ovchinnikova2017lexical}
I.~Ovchinnikova and A.~Pavlova, ``Lexical substitution and paraphasia in advanced dementia of the alzheimer type,'' \emph{Psychology of Language and Communication}, vol.~21, no.~1, pp. 306--324, 2017.

\bibitem{zolnoori2023adscreen}
M.~Zolnoori, A.~Zolnour, and M.~Topaz, ``Adscreen: A speech processing-based screening system for automatic identification of patients with alzheimer's disease and related dementia,'' \emph{Artificial Intelligence in Medicine}, vol. 143, p. 102624, 2023.

\bibitem{radford2023robust}
A.~Radford, J.~W. Kim, T.~Xu, G.~Brockman, C.~McLeavey, and I.~Sutskever, ``Robust speech recognition via large-scale weak supervision,'' in \emph{International conference on machine learning}.\hskip 1em plus 0.5em minus 0.4em\relax PMLR, 2023, pp. 28\,492--28\,518.

\bibitem{sanguedolce2024whisper}
G.~Sanguedolce, S.~Brook, D.~C. Gruia, P.~A. Naylor, and F.~Geranmayeh, ``When whisper listens to aphasia: Advancing robust post-stroke speech recognition,'' in \emph{Proc. Interspeech 2024}, 2024, pp. 1995--1999.

\bibitem{wang2022conformer}
T.~Wang, J.~Deng, M.~Geng, Z.~Ye, S.~Hu, Y.~Wang, M.~Cui, Z.~Jin, X.~Liu, and H.~Meng, ``Conformer based elderly speech recognition system for alzheimer's disease detection,'' in \emph{Proc. Interspeech 2022}, 2022.

\bibitem{10825055}
T.~Deußer, A.~M. Siddiqi, L.~Sparrenberg, T.~Adams, C.~Bauckhage, and R.~Sifa, ``Fusing speech and language models for dementia detection,'' in \emph{2024 IEEE International Conference on Big Data (BigData)}, 2024, pp. 3908--3914.

\bibitem{10448004}
J.~Li and W.-Q. Zhang, ``Whisper-based transfer learning for alzheimer disease classification: Leveraging speech segments with full transcripts as prompts,'' in \emph{ICASSP 2024 - 2024 IEEE International Conference on Acoustics, Speech and Signal Processing (ICASSP)}, 2024, pp. 11\,211--11\,215.

\bibitem{rohanian2021alzheimer}
M.~Rohanian, J.~Hough, and M.~Purver, ``Alzheimer's dementia recognition using acoustic, lexical, disfluency and speech pause features robust to noisy inputs,'' in \emph{Proc. Interspeech 2021}, 2021.

\bibitem{nasreen2021alzheimer}
S.~Nasreen, M.~Rohanian, J.~Hough, and M.~Purver, ``Alzheimer’s dementia recognition from spontaneous speech using disfluency and interactional features,'' \emph{Frontiers in Computer Science}, vol.~3, p. 640669, 2021.

\bibitem{soleimani2024impact}
R.~Soleimani, S.~Guo, K.~L. Haley, A.~Jacks, and E.~Lobaton, ``The impact of pause and filler word encoding on dementia detection with contrastive learning,'' \emph{Applied Sciences}, vol.~14, no.~19, p. 8879, 2024.

\bibitem{wagner2024crisperwhisper}
L.~Wagner, B.~Thallinger, and M.~Zusag, ``Crisperwhisper: Accurate timestamps on verbatim speech transcriptions,'' in \emph{Proc. Interspeech 2024}, 2024.

\bibitem{pitt}
\BIBentryALTinterwordspacing
J.~T. Becker, F.~Boiler, O.~L. Lopez, J.~Saxton, and K.~L. McGonigle, ``The natural history of alzheimer's disease: Description of study cohort and accuracy of diagnosis,'' \emph{Archives of Neurology}, vol.~51, no.~6, pp. 585--594, 06 1994. [Online]. Available: \url{https://doi.org/10.1001/archneur.1994.00540180063015}
\BIBentrySTDinterwordspacing

\bibitem{Kempler}
\BIBentryALTinterwordspacing
D.~Kempler, S.~Curtiss, and C.~Jackson, ``Syntactic preservation in alzheimer's disease,'' \emph{Journal of Speech, Language, and Hearing Research}, vol.~30, no.~3, pp. 343--350, 1987. [Online]. Available: \url{https://pubs.asha.org/doi/abs/10.1044/jshr.3003.343}
\BIBentrySTDinterwordspacing

\bibitem{cookie_theft}
C.~Roth, ``Boston diagnostic aphasia examination,'' \emph{Encyclopedia of clinical neuropsychology}, pp. 428--430, 2011.

\bibitem{macwhinney2014childes}
B.~MacWhinney, \emph{The CHILDES project: Tools for analyzing talk, Volume I: Transcription format and programs}.\hskip 1em plus 0.5em minus 0.4em\relax Psychology Press, 2014.

\bibitem{akinrintoyo2024co}
E.~Akinrintoyo and N.~Salomons, ``Co-designing icst with social robots for long-term in-home deployment for persons with dementia,'' \emph{Proceedings of the 2024 ACM/IEEE International Conference on Human-Robot Interaction Workshop on Ageing in Place}, 2024.

\bibitem{akinrinE}
------, ``In-home social robots design for cognitive stimulation therapy in dementia care,'' \emph{2025 34th IEEE International Conference on Robot and Human Interactive Communication (RO-MAN)}, 2025.

\bibitem{yates2015development}
L.~A. Yates, P.~Leung, V.~Orgeta, A.~Spector, and M.~Orrell, ``The development of individual cognitive stimulation therapy ({iCST}) for dementia,'' \emph{Clinical interventions in aging}, pp. 95--104, 2015.

\bibitem{srinivasan2019misty}
S.~Srinivasan, ``Misty-{A} development platform for socially assistive robots [student's corner],'' \emph{IEEE Robotics \& Automation Magazine}, vol.~26, no.~2, pp. 103--105, 2019.

\bibitem{jeronimo2018formal}
J.~Jer{\'o}nimo, T.~Queir{\'o}s, E.~Cheniaux, and D.~Telles-Correia, ``Formal thought disorders--historical roots,'' \emph{Frontiers in Psychiatry}, vol.~9, p. 572, 2018.

\bibitem{weiner2017manual}
J.~Weiner, M.~Engelbart, and T.~Schultz, ``Manual and automatic transcriptions in dementia detection from speech.'' in \emph{Interspeech}, 2017, pp. 3117--3121.

\bibitem{pan2020improving}
Y.~Pan, B.~Mirheidari, M.~Reuber, A.~Venneri, D.~Blackburn, and H.~Christensen, ``Improving detection of alzheimer’s disease using automatic speech recognition to identify high-quality segments for more robust feature extraction,'' in \emph{Proceedings of Interspeech 2020}.\hskip 1em plus 0.5em minus 0.4em\relax International Speech Communication Association (ISCA), 2020, pp. 4961--4965.

\bibitem{ye2021development}
Z.~Ye, S.~Hu, J.~Li, X.~Xie, M.~Geng, J.~Yu, J.~Xu, B.~Xue, S.~Liu, X.~Liu \emph{et~al.}, ``Development of the cuhk elderly speech recognition system for neurocognitive disorder detection using the dementiabank corpus,'' in \emph{ICASSP 2021-2021 IEEE International Conference on Acoustics, Speech and Signal Processing (ICASSP)}.\hskip 1em plus 0.5em minus 0.4em\relax IEEE, 2021, pp. 6433--6437.

\bibitem{pan2025two}
Y.~Pan, B.~Mirheidari, D.~Blackburn, and H.~Christensen, ``A two-step attention-based feature combination cross-attention system for speech-based dementia detection,'' \emph{IEEE Transactions on Audio, Speech and Language Processing}, 2025.

\end{thebibliography}

\end{document}